# Role of photon recycling in perovskite solar cells

M. Ryyan Khan [‡], Xufeng Wang [‡], Reza Asadpour, Mark Lundstrom, and Muhammad A. Alam [a)]

[1]*Electrical and Computer Engineering Department, Purdue University, West Lafayette, IN-47907, USA*

Nearly perfect photon recycling helped GaAs cells achieve the highest efficiency ever reported for a solar cell. Recent reports of photon recycling in perovskite solar cells suggest that, once optimized, it may as well achieve GaAs-like performance. In this paper, we show that GaAs and perovskite cells recycle photons in different ways. ***First,*** although bare-perovskite has been shown to have lifetimes (~1µs) in the radiative limit, non-radiative recombination at the transport layers restricts the solar cell operation far below the 'photon-recycling' regime. GaAs cells have no such limitation. ***Second***, even if the transport layers were optically and electrically perfect, the poor mobility of the perovskite layer would still restrict the optimum thickness ~1µm. Thus, a very high quality mirror (reflectivity >96%) is required to utilize photon-recycling. The mirror reflectivity restriction was far more relaxed for the thicker (~2 − 3µm) GaAs cells. Therefore, a nontrivial co-optimization of device geometry, mirror reflectivity, and material choice is necessary for achieving highest theoretical efficiency anticipated for perovskite cells.

## I. Introduction and background

There has been tremendous progress in perovskite solar cell performance since the 3.8% cell in 2009[1] -- the efficiency has now reached 22.1%. From this point onward, rapid development in performance will be difficult, and can only be achieved through meticulous material[2-9] and device engineering[10-13]. Earlier studies have predicted the performance limits of these solar cells [14-16]—but given the recent major improvements in material quality, these predictions should be re-visited.

The ultra-high efficiencies in GaAs cells are now near the radiative limit. In the radiative limit, photon recycling helps improve the open circuit voltage to very close to the Shockley-Quiesser (SQ) limit. Recent studies have reported very high bulk lifetime (in order of µs) in perovskites with very good external luminescence[17]. Therefore, perovskites show prospects for joining GaAs as an ultra-high performance solar cell[18].

At the current stage of perovskite PV technology, an obvious question arises: Can we follow the same design principles of GaAs solar cells to reach high performance in perovskite? To address this question, we need to determine processes that limit photon recycling in perovskites. For example, it is not clear how high the SRH lifetime needs to be for perovskites to reach the radiative limit. Parasitic optical losses and non-radiative recombination in the transport layers will limit the potential efficiency gain of these cells. There have been several works to minimize optical losses in transport layers (TLs). Reduced band-mismatch by engineering work-function and new materials have improved open circuit voltage ($V_{OC}$) and fill-factor (*FF*)[19-21].

In this paper, we lay out the design considerations towards an "ideal" perovskite solar cell. We use self-consistent opto-electronic simulation to analyze the role of absorption, photon recycling, and transport in the device. We first briefly describe this simulation framework. Photon recycling inside perovskite solar cells and the physics of external radiative efficiency (ERE) are then discussed, focusing on the effects of several critical design factors including SRH lifetime, absorber thickness, and mirror reflectivity. We present a breakdown on the loss components in a perovskite solar cell and explain the transition of perovskite cells from the non-radiative to the radiative limit. The principles for designing perovskite solar

---

[a)] Electronic mail: alam@purdue.edu.

[‡]These authors contributed equally

cells that operate near the SQ limit can be summarized as follows:

(i) Parasitic absorption in the TLs degrades both the short circuit current ($J_{SC}$) and the photon-recycling. TLs should be chosen to have weak or no absorption above the perovskite band-edge.
(ii) Although the bare-perovskites may have reached lifetimes close to the radiative limit, the TL/perovskite interface greatly enhances non-radiative recombination. This lowers the effective carrier lifetime in the device. Therefore, the solar cells cannot reach "GaAs-like" performance although the perovskite-layer itself has very high quality.
(iii) Finally, once the effective lifetime in the device has been improved close to the radiative limit, the poor mobility in perovskites still limits its optimal thickness to ~1μm. For photon-recycling in such a thin layer, the mirror reflectivity is required to be extremely high (>96%) for good photon-recycling.

## II. Simulation details

Simulations in this work are based on techniques developed earlier for exploring the design of GaAs solar cells that operate near the Shockley-Queisser (SQ) limit [22-26]. In this technique, radiative recombination and photon recycling are rigorously calculated. First, the intrinsic radiative recombination rate is obtained from the van Roosbroeck-Shockley equation as a property of the perovskite material without dependence on geometry. Photons emitted from this intrinsic radiative recombination are isotopically partitioned into 0.2-degree resolution covering all directions are ray-traced around the solar cell. The ray tracing module and then uses Beer-Lambert equation for absorption and Fresnel equations at material boundaries. During the recycling process, photons may be emitted out of the cell as photoluminescence via the escape cone, parasitically absorbed by other non-active layers or the backside mirror, or re-absorbed by perovskite to generate new electron-hole pairs. For each ray, the tracing ends when less than 0.01% of original intensity remains. This photon recycling optical calculation is coupled with semiconductor electronic transport equations to form a self-consistent and electro-optically coupled framework in the numerical simulator Sentaurus [27].

We consider a typical perovskite cell structure, which consists of a layer of perovskite (~300 nm) sandwiched between an Electron Transport Layer (ETL) and a Hole Transport Layer (HTL). We will collectively refer to the ETL and HTL as Transport Layers (TLs). The key parameters used are listed in Table S1. SRH recombination is modeled using a single, "effective" value, without distinguishing between surface and bulk SRH components. It can be interpreted as the effective lifetime of the combined HTL/perovskite/ETL structure measured directly through, for example, Time Resolved Photoluminescence (TRPL). Although we numerically study the inverted configuration, the "effective lifetime" approach allows us to generalize our conclusions to both the standard (e.g., ITO/TiO$_2$ (ETL)/perovskite/HTL) and an inverted (e.g., ITO/PEDOT (HTL)/perovskite/ETL) device.

## III. Understanding photon-recycling in perovskite cells

One may trivially approach the SQ limit by removing all SRH recombination, leaving only the radiative and Auger losses that fundamentally exist. In this limit one source of degradation in $J_{SC}$ is an imperfect mirror. Due to *conventionally used* 300 nm thin perovskite layer, the sunlight is not absorbed in a single pass—the mirror is therefore important in reflecting the light to provide effectively a longer absorption length. Thus, as seen in Fig. 1(a), absorption in the perovskite, i.e., $J_{SC}$ will decrease with mirror reflectivity $R_{mirr}$. A moderate $R_{mirr}$~80% lowers $J_{SC}$ by ~0.5 mA/cm$^2$ compared to ideal. Parasitic absorption in TLs can further reduce $J_{SC}$ (see supplement). Within an inverted cell, light enters through the TCO, then travels through HTL and reaches the perovskite.

Let us now consider photon recycling. Photon recycling causes the internally emitted photons to be re-absorbed within the perovskite layer. In a planar structure, a fraction ($\sim 1/n^2$, where $n$ is the refractive index) of the internal emission escapes and radiates outside the device. This contributes to the fundamental gap between $E_G$ and $V_{OC}$. The internally emitted photons outside the escape cone go through several bounces ($\sim 1/2\alpha L$, with $\alpha$ being the absorption coefficient) inside the perovskite layer before being re-absorbed.

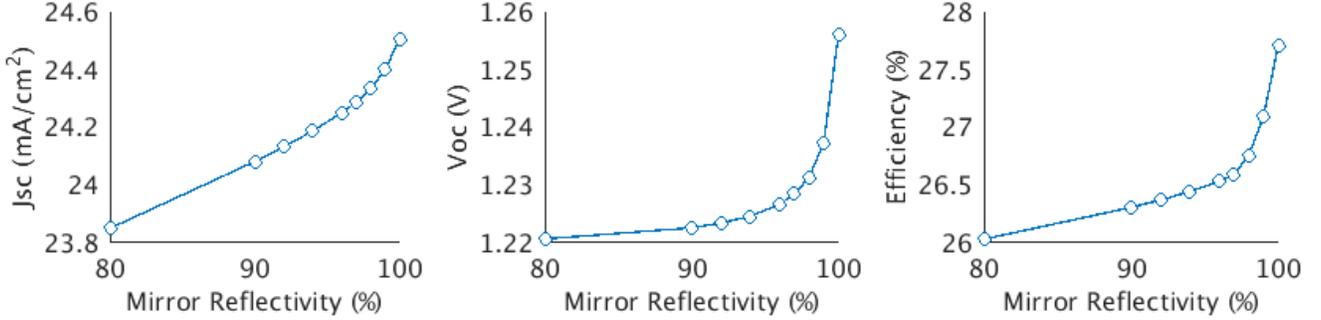

Fig. 1: (a) Jsc, (b) Voc, and (c) efficiency as function of mirror reflectivity assuming no optical losses in TLs. No SRH recombination is included in these results.

With no optical loss in the TLs, most of the reduction in $V_{OC}$ occurs as $R_{mirr}$ drops from 100% to 95%. This is because majority of the internally emitted photons are lost to the imperfect mirror thereby lowering photon recycling. The reason for such sharp decrease is the thin cell thickness of 300 nm. In a thinner cell, the photon requires more number of bounces off the mirror to travel its absorption length. Therefore, for the same mirror reflectivity, a thinner cell will have more radiative recombination loss due to parasitic mirror absorption.

Even with no parasitic optical losses ($R_{mirr} = 100\%$), as seen in Fig. 1(c), $\eta \approx 27.7\%$ in the radiative limit, which is lower than the SQ limit (~33%). This is due to two reasons: (i) low mobility, and (ii) incomplete absorption in the thin 300 nm perovskite. On the first reason, it is known that low mobility can lower the radiative limit below the SQ value [28]. On the second reason, a thicker perovskite layer is required to completely absorb the above bandgap photons. Unfortunately, SRH recombination in a thick perovskite competes with the photo-absorption— yielding an optimum thickness $L$ for a given SRH lifetime ($\tau_{SRH}$), as follows.

## IV. Role of SRH recombination

### A. Competition between radiative and non-radiative recombinations: physics of ERE

Radiative efficiency is an important figure of merit for photon management. The External Radiative Efficiency (ERE) is defined by ratio of emission and $J_{SC}$. PL measurements can easily determine ELE, and therefore the metric has been used to access the cell quality, and compare cell performances across different technologies. In Fig. 2(a), ERE is shown for both 80% (solid line) and 100% (dashed line) mirror reflectivity along the maximum efficiency peak. The degradation in photon recycling from 100% to 80% $R_{mirr}$ reduces ERE significantly. Contribution of radiative recombination increases, as seen by the increase in $\eta_{ext}$ with $\tau_{SRH}$. We observe that $\eta_{ext}$ saturates for $\tau_{SRH} \sim 1$ μs (5 μs) with $R_{mirr} = 80\%$ (100%). With $\tau_{SRH} \sim 1$ μs, ERE $\eta_{ext}$ reaches ~5% for an 80% mirror. To compare, the best GaAs cells have ERE of ~20%. The overall ERE for 80% mirror, shown in Fig. 2(b), is no more than 8%.

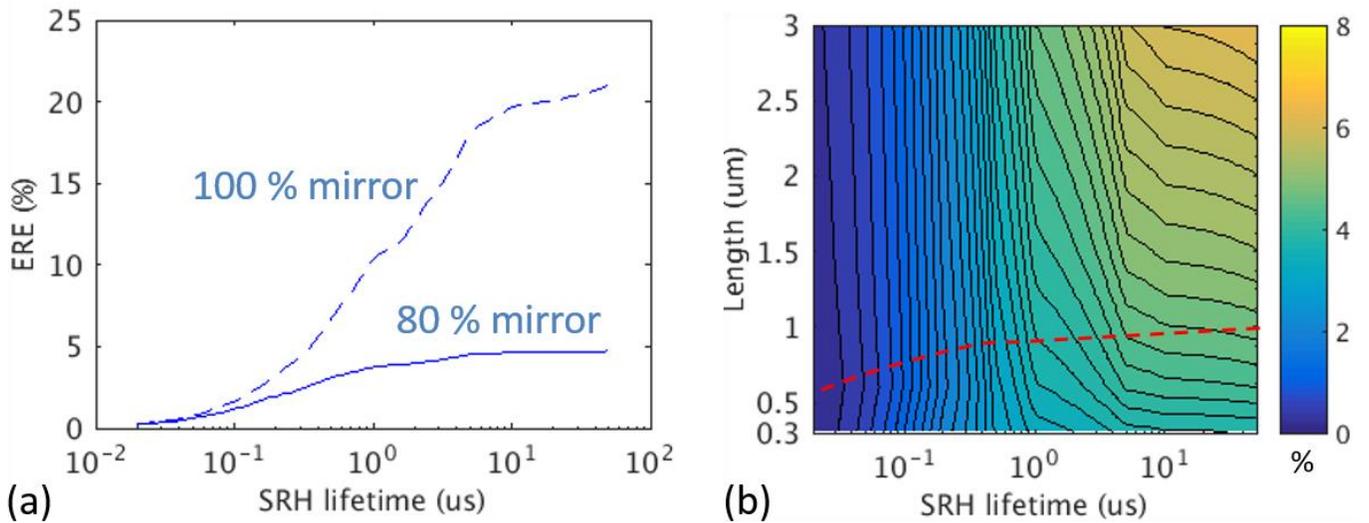

Fig. 2: (a) ERE vs. $\tau_{SRH}$ for 100% (dashed line) and 80% (solid line) mirror reflectivity along the maximum efficiency peak. (b) ERE vs. $\tau_{SRH}$ and $L$ for 80% mirror reflectivity.

As we will discuss later, there is an optimum perovskite thickness $L$ for each $\tau_{SRH}$ which maximizes the efficiency, as shown by the red dashed line in Fig. 3(d). Interestingly, at high $\tau_{SRH}$ region, ERE increases with higher $L$, while cell efficiency shows an optimal $L$ for each $\tau_{SRH}$. The difference is attributed to two factors. First, recall that ERE is defined by the ratio between emission and $J_{SC}$. In Fig. 3(a), $J_{SC}$ decreases for absorbers thicker than 1 um, and as a result, even though the cell is less efficient, the ERE increases by definition. This is usually not an issue in cells with high mobility materials, such as silicon and GaAs. The second factor is the change of radiative and non-radiative recombination ratio. The dominating non-radiative recombination mechanism at high $\tau_{SRH}$ region is Auger recombination, which has a cubic dependence with carrier concentration, Fig. 4(b). As the $V_{OC}$ decreases with increasing $L$, the ratio of the radiative to Auger recombination increases slightly. Combining these two factors (Fig. 2(b) vs. 3(d)), we conclude that optimizing the ERE may not optimize the cell efficiency. Therefore, ERE should not be taken as the exclusive design metric for high-efficiency perovskite solar cells.

## B. Optimal design of the perovskite solar cell

Having presented a general analysis of photon recycling and ERE in perovskite solar cells, we now focus on the cell efficiency and explore the key question: how does one improve today's cell toward the SQ limit? The performance parameters of the perovskite solar cell strongly co-depend on the active layer thickness $L$, and the effective lifetime ($\tau_{SRH}$). We now map out the effects of these key parameters and design principles toward ideal perovskite solar cells.

The design space for solar cells with a modest mirror reflectivity of 80% (common metal reflectors) is shown collectively in Fig. 3. As expected, both $J_{SC}$ (Fig. 3(a)) and $V_{OC}$ (Fig. 3(c)) improve with increasing $\tau_{SRH}$. The trend with varying length ($L$) is, however, different. Increasing length lowers $V_{OC}$ due to higher SRH recombination within the increased perovskite volume. This is more evident with lower SRH lifetime. On the other hand, photo-absorption will increase and saturate with $L$ as shown in Fig. 3(b) (red line labeled "J$_{gen}$"). Due to the low mobility of perovskite, longer $L$ and lower SRH lifetime decrease the collection of these light-generated carriers. As a result of this absorption-collection tradeoff, the $J_{SC}$ is optimized at a finite thickness.

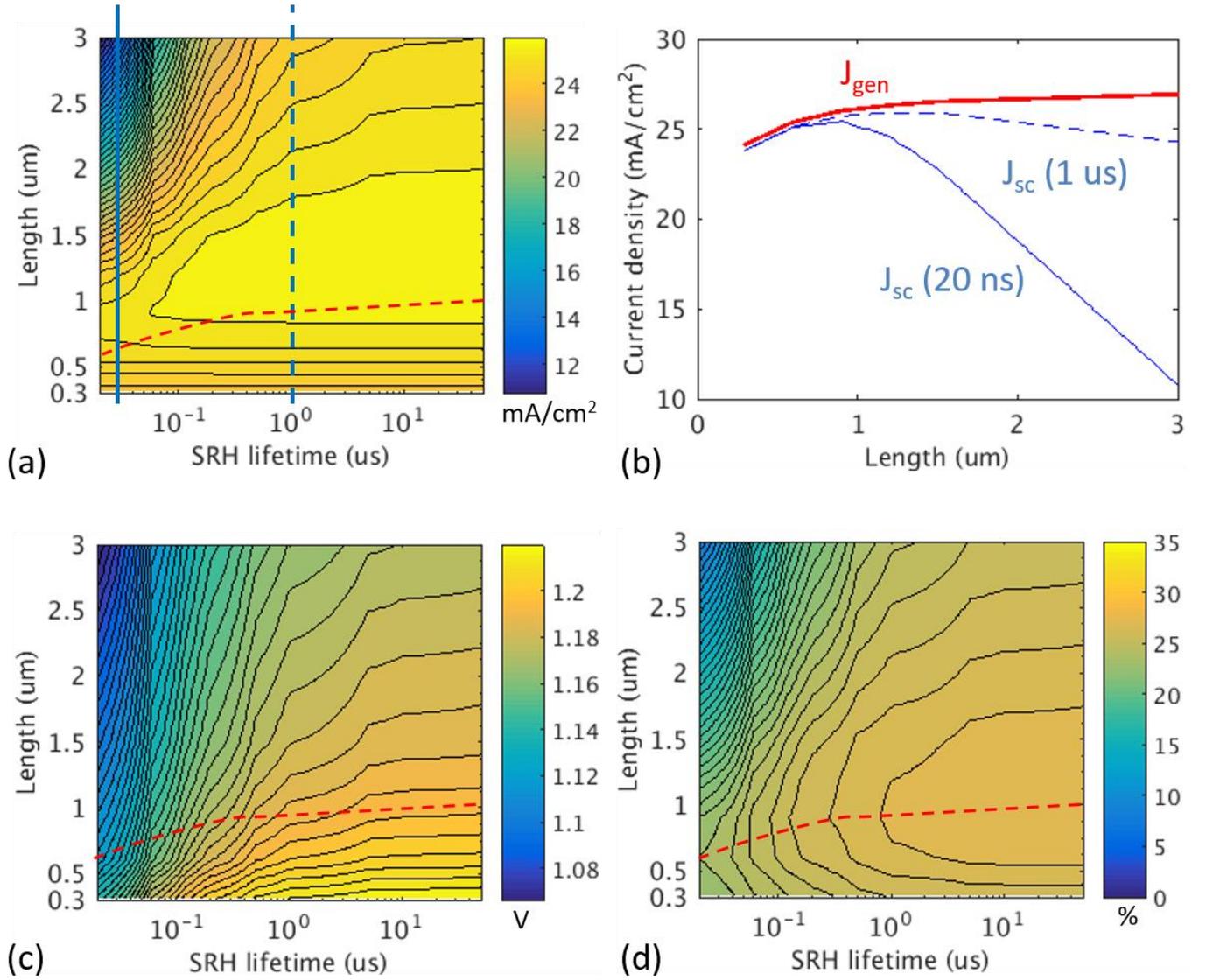

Fig. 3: Performance maps for solar cells with 80% mirror reflectivity. (a) Jsc vs. $\tau_{NR}$ and $L$. (b) Jsc vs. L taken at 20 ns (solid blue line) and 1 us (dashed blue line), marked by solid and dashed blue lines in (a) respectively. (c) Voc vs. $\tau_{NR}$ and $L$. (d) Efficiency vs. $\tau_{NR}$ and $L$. For each of the contour plot, the maximum efficiency at each $\tau_{NR}$ is shown as the red dashed line.

Overall, the tradeoffs among $L$, $J_{SC}$ and $V_{OC}$ result in an optimum thickness for maximum $\eta$, as shown in Fig. 3(d). This "maximum $\eta$ peak line" at each $\tau_{SRH}$ is shown as the red dashed line in Fig. 3(a), (c), and (d), and it provides design guidance for optimizing device geometry with improving perovskite material quality. For each $\tau_{SRH}$, the $\eta$ maximum occurs at lower $L$ than that of $J_{SC}$. This is because $V_{OC}$ always increases toward lower $L$. Moving along the x-axis, as $\tau_{SRH}$ increases, the cell benefits more from having higher $J_{SC}$ with increased $L$. This increase in $L$ eventually saturates to ~1 um, due to the SRH no longer being the dominant recombination mechanism. The comparison among various loss mechanisms will be discussed shortly.

Focusing on this maximum $\eta$ peak line, Fig. 4(a) shows the maximum $\eta$ vs. $\tau_{SRH}$ for 80% (solid line) and 100% (dashed line) mirror reflectivity. Efficiency with imperfect mirror of $R_{mirr} = 80\%$ is always lower than with a perfect mirror as expected. This is due to the combined loss in photo-absorption ($J_{SC}$) and photon recycling ($V_{OC}$). With an ideal mirror and high $\tau_{SRH}$, one reaches a maximum $\eta$ of 28.8%. This is however significantly lower than the SQ limit at ~33% predicted for perovskite. This difference comes from the Auger recombination and low mobility—two practical factors not considered in SQ limit.

Let us inspect the loss components in the state-of-the-art perovskite solar cells. In Fig. 4(b), we show loss components vs. $\tau_{SRH}$ along the maximum efficiency peak for 80% mirror reflectivity. With consideration of the finite mobility and Auger recombination, we predict an upper efficiency limit for perovskite solar cell at 28.8% (see dashed line in Fig. 5(a)). There are five losses involved for any cell below this limit: (1) incomplete absorption (black area labeled as "In-abs"), (2) Auger recombination (green area labeled as "Auger"), (3) rear mirror loss (red area labeled as "Mirror"), (4) SRH recombination (blue area labeled as "SRH"), and (5) photoluminescence emission (orange area labeled as "PL"). As expected in the low $\tau_{SRH}$ region, SRH recombination dominates. The efficiency loss due to incomplete absorption is also significant due to the optimal thickness at this region being < 1 um (see Fig. 3(d)). As $\tau_{SRH}$ increases, radiative recombination begins to dominate over SRH recombination. Photon-recycling plays a more prominent role in the device performance for higher $\tau_{SRH}$. Therefore, loss due to rear mirror becomes significant for $\tau_{SRH} > 1$ μs. This efficiency loss stabilizes at ~1%, since the optimal thickness stabilizes at ~ 1 um. Auger recombination is ~10% of the mirror loss. In our numerical model and the loss-analysis study, we have not considered loss due to reflection of light from the top surface. Approximately 15% top-surface reflection of the incident light can be estimated from ~85% EQE[8, 19].

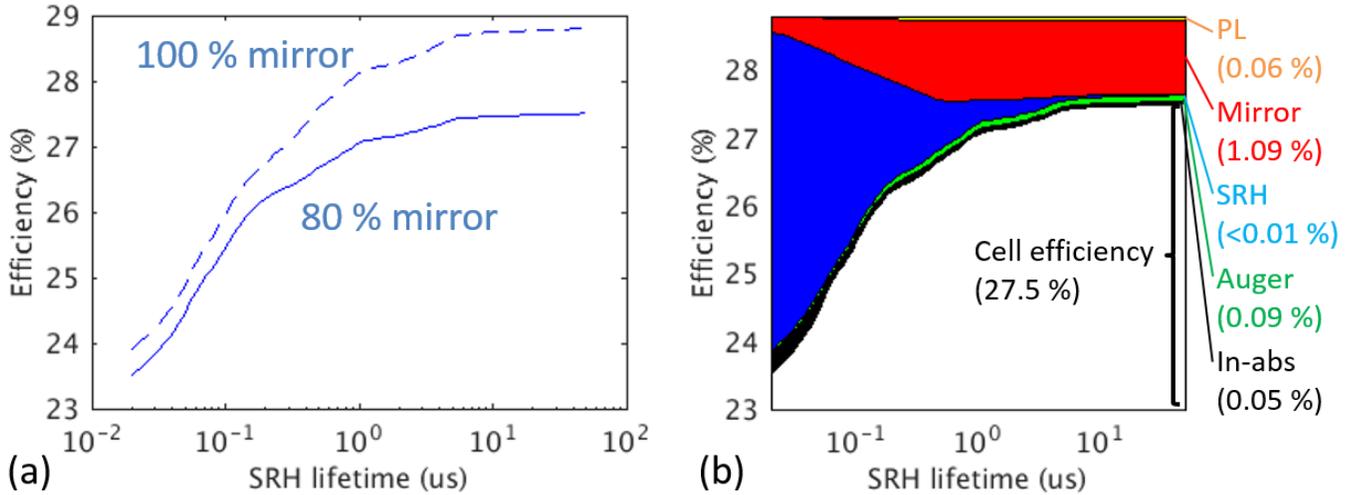

Fig. 4: (a) Efficiency for 100% (dashed line) and 80% (solid line) mirror reflectivity. (b) Efficiency loss for each loss components vs. $\tau_{NR}$ for 80% mirror reflectivity. The efficiency maximum is at 28.8%. At the right side of the plot, efficiency losses for each loss component with SRH lifetime of 50 us are shown.

This work has combined the electronic losses in perovskites and TLs into a single effective lifetime $\tau_{SRH}$. In practice, TLs are known to be a significant challenge to design and optimize. Although the perovskite itself may have high quality, with recent works showing that intrinsic perovskite SRH lifetime can reach ~1 μs, the TLs and their interfaces can be defective and introduce significant SRH recombination. Our analysis with 1μs perovskite lifetime but poor interface (in supplementary materials) show how the performance is restricted by the TLs alone. These results are consistent with effective $\tau_{SRH} < 50$ ns. Therefore, improving the perovskite alone cannot make these devices more efficient: increasing the effective lifetime by interface engineering and reducing recombination in the transport layers is a key factor.

## V. Conclusions

Through self-consistent, electro-optical coupled device simulation, we showed that the design of perovskite solar cells that operate near the Shockley-Queisser limit is very different from that of GaAs. In GaAs, the design aspect is to increase the photon recycling, resulting in many research heavily focused on the mirror quality and novel photon management schemes. The situation is more complicated in perovskite solar cells.

*First*, the TLs must be selected to have very low absorption of photons at the perovskite bandgap energy. This is important for immediate improvement in $J_{SC}$ and for allowing better photo-recycling once electronic losses are suppressed.

*Second*, interface and bulk SRH recombination must be lessened to increase $\tau_{SRH}$ and gain access to the higher efficiency part of the mapping. Several works in the current literature present high quality perovskites with lifetime on the order of 1μs. However, we stress once again that the effective lifetime of the full HTL/perovskite/ETL structure is what defines the performance of the solar cell. Therefore, lifetimes for both bare-perovskite and HTL/perovskite/ETL structures should be analyzed by TRPL experiments. As we have shown, the efficiency of the cell saturates beyond effective lifetime $\tau_{SRH} > 1$μs. Therefore, perovskites may have reached the quality required to reach its fundamental performance limit—the major hurdle is to improve the mobility of TLs and create better interfaces.

*Finally*, as the TLs are improved, the effective-$\tau_{SRH}$ would dictate the optimum perovskite thickness $L$. Also, due to the poor mobility of perovskites, the optimal thickness of the cell is limited to ~1 um, which causes the mirror reflectivity to strongly affect photon recycling. Without an advanced mirror (> 96 %), enhanced photon-recycling and efficiency gain from mirror improvement will not be significant.

These unique challenges must be addressed in order for perovskite solar cells to reach toward the fundamental, radiative limit.

**Acknowledgement**

We gratefully acknowledge research support from the NCN NEEDS program, which is funded by the National Science Foundation, contract 1227020-EEC, and the Semiconductor Research Corporation. This work was also supported in part by the Solar Energy Research Institute for India and the United States (SERIIUS), funded jointly by the U.S. Department of Energy under Subcontract DE-AC36-08GO28308 to the National Renewable Energy Laboratory, Golden, CO, USA) and the Government of India, through the Department of Science and Technology under Subcontract IUSSTF/JCERDC-SERIIUS/2012.

**REFERENCES**

[1] A. Kojima, K. Teshima, Y. Shirai, and T. Miyasaka, "Organometal halide perovskites as visible-light sensitizers for photovoltaic cells," *J Am Chem Soc,* vol. 131, pp. 6050-1, May 6 2009.

[2] J. H. Heo, S. H. Im, J. H. Noh, T. N. Mandal, C.-S. Lim, J. A. Chang*, et al.*, "Efficient inorganic–organic hybrid heterojunction solar cells containing perovskite compound and polymeric hole conductors," *Nature Photonics,* vol. 7, pp. 486-491, 2013.

[3] J. Burschka, N. Pellet, S. J. Moon, R. Humphry-Baker, P. Gao, M. K. Nazeeruddin*, et al.*, "Sequential deposition as a route to high-performance perovskite-sensitized solar cells," *Nature,* vol. 499, pp. 316-9, Jul 18 2013.

[4] J. H. Noh, S. H. Im, J. H. Heo, T. N. Mandal, and S. I. Seok, "Chemical management for colorful, efficient, and stable inorganic-organic hybrid nanostructured solar cells," *Nano Lett,* vol. 13, pp. 1764-9, Apr 10 2013.

[5] W. Nie, H. Tsai, R. Asadpour, J. C. Blancon, A. J. Neukirch, G. Gupta*, et al.*, "Solar cells. High-efficiency solution-processed perovskite solar cells with millimeter-scale grains," *Science,* vol. 347, pp. 522-5, Jan 30 2015.

[6] Y. Y. Zhou, M. J. Yang, W. W. Wu, A. L. Vasiliev, K. Zhu, and N. P. Padture, "Room-temperature crystallization of hybrid-perovskite thin films via solvent-solvent extraction for high-performance solar cells," *Journal of Materials Chemistry A,* vol. 3, pp. 8178-8184, 2015.

[7] W. S. Yang, J. H. Noh, N. J. Jeon, Y. C. Kim, S. Ryu, J. Seo*, et al.*, "High-performance photovoltaic perovskite layers fabricated through intramolecular exchange," *Science,* vol. 348, pp. 1234-7, Jun 12 2015.

[8] N. J. Jeon, J. H. Noh, W. S. Yang, Y. C. Kim, S. Ryu, J. Seo*, et al.*, "Compositional engineering of perovskite materials for high-performance solar cells," *Nature,* vol. 517, pp. 476-80, Jan 22 2015.

[9] M. Saliba, T. Matsui, J. Y. Seo, K. Domanski, J. P. Correa-Baena, M. K. Nazeeruddin*, et al.*, "Cesium-containing triple cation perovskite solar cells: improved stability, reproducibility and high efficiency," *Energy Environ Sci,* vol. 9, pp. 1989-1997, Jun 8 2016.

[10] M. Liu, M. B. Johnston, and H. J. Snaith, "Efficient planar heterojunction perovskite solar cells by vapour deposition," *Nature,* vol. 501, pp. 395-8, Sep 19 2013.

[11] J. You, Z. Hong, Y. M. Yang, Q. Chen, M. Cai, T. B. Song*, et al.*, "Low-temperature solution-processed perovskite solar cells with high efficiency and flexibility," *ACS Nano,* vol. 8, pp. 1674-80, Feb 25 2014.

[12] O. Malinkiewicz, A. Yella, Y. H. Lee, G. M. Espallargas, M. Graetzel, M. K. Nazeeruddin*, et al.*, "Perovskite solar cells employing organic charge-transport layers," *Nature Photonics,* vol. 8, pp. 128-132, 2013.

[13] A. R. Pascoe, M. Yang, N. Kopidakis, K. Zhu, M. O. Reese, G. Rumbles*, et al.*, "Planar versus mesoscopic perovskite microstructures: The influence of CH3NH3PbI3 morphology on charge transport and recombination dynamics," *Nano Energy,* vol. 22, pp. 439-452, 2016.

[14] W. E. I. Sha, X. Ren, L. Chen, and W. C. H. Choy, "The efficiency limit of CH3NH3PbI3 perovskite solar cells," *Applied Physics Letters,* vol. 106, p. 221104, 2015.

[15] S. Agarwal and P. R. Nair, "Device engineering of perovskite solar cells to achieve near ideal efficiency," *Applied Physics Letters,* vol. 107, p. 123901, 2015.

[16] V. Gonzalez-Pedro, E. J. Juarez-Perez, W. S. Arsyad, E. M. Barea, F. Fabregat-Santiago, I. Mora-Sero*, et al.*, "General working principles of CH3NH3PbX3 perovskite solar cells," *Nano Lett,* vol. 14, pp. 888-93, Feb 12 2014.

[17] L. M. Pazos-Outon, M. Szumilo, R. Lamboll, J. M. Richter, M. Crespo-Quesada, M. Abdi-Jalebi*, et al.*, "Photon recycling in lead iodide perovskite solar cells," *Science,* vol. 351, pp. 1430-3, Mar 25 2016.

[18] E. Yablonovitch, "Lead halides join the top optoelectronic league," *Science,* vol. 351, p. 1401, Mar 25 2016.

[19] M. Saliba, S. Orlandi, T. Matsui, S. Aghazada, M. Cavazzini, J.-P. Correa-Baena*, et al.*, "A molecularly engineered hole-transporting material for efficient perovskite solar cells," *Nature Energy,* vol. 1, p. 15017, 2016.

[20] H. Zhou, Q. Chen, G. Li, S. Luo, T. B. Song, H. S. Duan*, et al.*, "Photovoltaics. Interface engineering of highly efficient perovskite solar cells," *Science,* vol. 345, pp. 542-6, Aug 1 2014.

[21] R. Fan, Y. Huang, L. Wang, L. Li, G. Zheng, and H. Zhou, "The Progress of Interface Design in Perovskite-Based Solar Cells," *Advanced Energy Materials,* p. 1600460, 2016.

[22] X. Wang, M. R. Khan, J. L. Gray, M. A. Alam, and M. S. Lundstrom, "Design of GaAs Solar Cells Operating Close to the Shockley–Queisser Limit," *IEEE Journal of Photovoltaics,* vol. 3, pp. 737-744, 2013.

[23] X. Wang, J. Bhosale, J. Moore, R. Kapadia, P. Bermel, A. Javey*, et al.*, "Photovoltaic Material Characterization with Steady-State and Transient Photoluminesence," ed, 2014.

[24] X. Wang, M. R. Khan, M. A. Alam, and M. Lundstrom, "Approaching the Shockley-Queisser limit in GaAs solar cells," in *IEEE Photovoltaic Specialists Conference*, 2012, pp. 002117-002121.

[25] X. Wang, M. R. Khan, M. Lundstrom, and P. Bermel, "Performance-limiting factors for GaAs-based single nanowire photovoltaics," *Optics Express,* vol. 22, p. A344, 2014.
[26] X. Wang and M. S. Lundstrom, "On the Use of Rau's Reciprocity to Deduce External Radiative Efficiency in Solar Cells," *IEEE Journal of Photovoltaics,* pp. 1-6, 2013.
[27] "Synopsys Sentaurus Semiconductor TCAD Software," ed. East Middlefield Road, Mountain View, CA 94043 USA.
[28] J. Mattheis, J. H. Werner, and U. Rau, "Finite mobility effects on the radiative efficiency limit of pn-junction solar cells," *Physical Review B,* vol. 77, Feb 2008.